\input harvmac
\input amssym.tex

\def\wt{\widetilde{\lambda}}
\def\P{\Bbb{P}}

\def\C{\Bbb{C}}

\lref\WittenNN{
E.~Witten,
``Perturbative gauge theory as a string theory in twistor space,''
arXiv:hep-th/0312171.
}

\lref\ParkeGB{
S.~J.~Parke and T.~R.~Taylor,
``An Amplitude For $N$ Gluon Scattering,''
Phys.\ Rev.\ Lett.\  {\bf 56}, 2459 (1986).
}

\lref\BerendsME{
F.~A.~Berends and W.~T.~Giele,
``Recursive Calculations For Processes With $N$ Gluons,''
Nucl.\ Phys.\ B {\bf 306}, 759 (1988).
}

\lref\NairBQ{
V.~P.~Nair,
``A Current Algebra For Some Gauge Theory Amplitudes,''
Phys.\ Lett.\ B {\bf 214}, 215 (1988).
}

\lref\PenroseWN{
R.~Penrose,
``Twistor Algebra,''
J.\ Math.\ Phys.\  {\bf 8}, 345 (1967);
R.~Penrose,
``Twistor Quantisation and Curved Space-Time,''
Int.~J.~Theor.~Phys.~{\bf 1} 61 (1968).
}

\lref\DixonWI{
L.~J.~Dixon,
``Calculating scattering amplitudes efficiently,''
arXiv:hep-ph/9601359.
}

\lref\KosowerXY{
D.~A.~Kosower,
``Light Cone Recurrence Relations For QCD Amplitudes,''
Nucl.\ Phys.\ B {\bf 335}, 23 (1990).
}

\lref\ChalmersUI{
G.~Chalmers and W.~Siegel,
``Simplifying algebra in Feynman graphs. I: Spinors,''
Phys.\ Rev.\ D {\bf 59}, 045012 (1999)
[arXiv:hep-ph/9708251];
G.~Chalmers and W.~Siegel,
``Simplifying algebra in Feynman graphs. II: Spinor helicity from the
spacecone,''
Phys.\ Rev.\ D {\bf 59}, 045013 (1999)
[arXiv:hep-ph/9801220].
}

\lref\BernMQ{
Z.~Bern, L.~J.~Dixon and D.~A.~Kosower,
``One loop corrections to five gluon amplitudes,''
Phys.\ Rev.\ Lett.\  {\bf 70}, 2677 (1993)
[arXiv:hep-ph/9302280].
}

\lref\BernQK{
Z.~Bern, G.~Chalmers, L.~J.~Dixon and D.~A.~Kosower,
``One loop N gluon amplitudes with maximal helicity violation via collinear
limits,''
Phys.\ Rev.\ Lett.\  {\bf 72}, 2134 (1994)
[arXiv:hep-ph/9312333];
G.~Mahlon,
``Multi-gluon helicity amplitudes involving a quark loop,''
Phys.\ Rev.\ D {\bf 49}, 4438 (1994)
[arXiv:hep-ph/9312276].
}

\lref\BernZX{
Z.~Bern, L.~J.~Dixon, D.~C.~Dunbar and D.~A.~Kosower,
``One Loop N Point Gauge Theory Amplitudes, Unitarity And Collinear Limits,''
Nucl.\ Phys.\ B {\bf 425}, 217 (1994)
[arXiv:hep-ph/9403226];

Z.~Bern, L.~J.~Dixon, D.~C.~Dunbar and D.~A.~Kosower,
``Fusing gauge theory tree amplitudes into loop amplitudes,''
Nucl.\ Phys.\ B {\bf 435}, 59 (1995)
[arXiv:hep-ph/9409265].
}

\lref\WittenCS{
E.~Witten,
``Chern-Simons gauge theory as a string theory,''
Prog.\ Math.\  {\bf 133}, 637 (1995)
[arXiv:hep-th/9207094].
}

\lref\ManganoXK{
M.~L.~Mangano, S.~J.~Parke and Z.~Xu,
``Duality And Multi-Gluon Scattering,''
Nucl.\ Phys.\ B {\bf 298}, 653 (1988).
}

\Title{\vbox{\baselineskip12pt
	\hbox{hep-th/0402016}
          \hbox{NSF-KITP-04-15}
}}{A Googly Amplitude from the B-model in Twistor Space}

\centerline{
Radu Roiban${}^\dagger$,
Marcus Spradlin${}^\ddagger$ and
Anastasia Volovich${}^\ddagger$}

\bigskip
\bigskip

\centerline{${}^\dagger$Department of Physics, University of California}
\centerline{Santa Barbara, CA 93106 USA}

\smallskip

\centerline{${}^\ddagger$Kavli Institute for Theoretical Physics}
\centerline{Santa Barbara, CA 93106 USA}

\bigskip
\bigskip

\centerline{\bf Abstract}

\bigskip
Recently it has been proposed that gluon scattering amplitudes
in gauge theory
can be computed from the D-instanton expansion of the
topological B-model on
$\P^{3|4}$, although only maximally helicity
violating (MHV) amplitudes have so far been obtained
from a direct B-model calculation.
In this note we compute the simplest non-MHV
gluon amplitudes ($++---$ and
$+-+--$) from the B-model as an integral over
the moduli space of degree 2 curves in $\P^{3|4}$
and
find perfect agreement with Yang-Mills theory.

\Date{February 2004}

\listtoc
\writetoc

\newsec{Introduction}

The relation between gauge theory and string theory is
one of the most important themes in modern theoretical physics.
Significant progress in understanding this relation has emerged from
the AdS/CFT correspondence, but this formulation does not allow direct
access to the perturbative states of the gauge theory,
making it difficult to calculate
experimentally relevant Yang-Mills (YM) scattering amplitudes.

Recently, Witten \WittenNN\ described a remarkable construction which
formulates 
${\cal N}=4$ supersymmetric Yang-Mills (SYM) theory as a full-fledged
string theory, the B-twisted topological string on $\P^{3|4}$. 
It was conjectured that the perturbative amplitudes of YM theory are 
recovered from instantonic D1-string calculations in this topological 
string theory. 
The origin of this proposal was the observation that YM
scattering amplitudes have unexpected properties
which seem to cry out for some deeper explanation. For
example, it has been known since
the work of Nair \NairBQ\ that maximally
helicity violating (MHV) tree-level amplitudes can be written
in terms of correlation functions of free fermionic currents on a 
2-sphere.
In \refs{\WittenNN}
it was shown, by considering a large number of examples, 
that more complicated YM amplitudes satisfy a number of
highly nontrivial differential identities.
These identities express the fact that YM scattering
amplitudes, when transformed to the twistor space
\PenroseWN\
 $\P^3$ of  Minkowski
space, 
are supported on curves whose genus and degree are
related to the number of
YM loops and to the number and helicities of the external legs.
The topological B-model was proposed as a candidate
theory which would 
expose these remarkable properties.

This proposal was used in \WittenNN\ to provide a context
for the calculation of \NairBQ\ and to
reproduce the MHV gluon scattering
amplitudes from the topological B-model by integrating
a certain free-fermion correlation function over the moduli
space of degree one curves in $\P^{3|4}$.
However, the question of whether the conjecture
might prove computationally useful for more complicated
amplitudes was  left open.

On the gauge theory side there exists a wealth of  results
on YM scattering amplitudes,
derived through a variety of methods. The complexity of the result
grows substantially as  additional negative helicity gluons are
added. Tree level MHV amplitudes (and their conjugates) with an
arbitrary number of external legs have been computed in
\refs{\ParkeGB,\ManganoXK}.
Powerful recurrence relations constructed in \BerendsME\ 
were exploited in \KosowerXY\ to calculate certain
amplitudes with three
negative helicity and arbitrarily
many positive helicity gluons.
General algorithms simplifying the calculations for
theories with massless particles have been devised in \ChalmersUI.
Substantial progress was achieved also in the calculation of loop
amplitudes by the use of string-inspired methods \BernMQ\ and
the exploitation of the constraints coming from collinear 
limits and unitarity \refs{\BernQK, \BernZX}.

In this note we provide strong further evidence for the conjecture
of \WittenNN\ by recovering the gauge theory 5-point amplitude
with three negative and two positive helicity\foot{Amplitudes with two
positive helicity gluons and an arbitrary number of negative helicity
ones
were called `googly' MHV amplitudes in \WittenNN.}
gluons ($++---$ and $+-+--$) from a
D-instanton computation in the open string
field theory of the B-model on $\P^{3|4}$.
These amplitudes are quite simple in gauge theory, since in
Lorentzian signature they are complex conjugates of MHV amplitudes.
However the B-model calculation involves an apparently quite nontrivial
integral over the moduli space of degree 2 curves in $\P^{3|4}$.

It may appear that our calculation amounts to using an
elephant gun to shoot a fly.  We are optimistic however
that the calculational techniques employed in this paper
will generalize, with
some refinement,  to more complicated amplitudes.
In particular, since we
now know that it is possible
to evaluate an integral over the moduli
space of degree 2 curves, it does not appear exceedingly
difficult to add an arbitrary number of additional positive
helicity gluons (which do not change the degree).
The corresponding gauge theory amplitudes, in the few
cases which are known, are rather complicated \KosowerXY.
Ultimately the correspondence between YM
theory and the topological B-model
may provide powerful new insights as well as
concrete calculational tools
for studying gauge theory amplitudes.

We begin in \S 2 with a review of the gauge theory results and
of the observation that they are supported on certain classes of
curves in the twistor space of Minkowski space and proceed in \S 3 to
briefly review the string theory construction. We then use the 
general expression for the scattering amplitudes described in that
section to recover in \S 4 the gauge theory result for the 5-point
amplitude.

\newsec{Helicity Amplitudes in Gauge Theory}

We consider tree-level scattering amplitudes ${\cal{A}}_n$ of
$n$ gluons in YM theory.  At tree level, the amplitudes
do not depend on the presence or absence of supersymmetry.
All formulae will be written in a manifestly ${\cal{N}}=4$
supersymmetric way, but the resulting gluon amplitudes are equally valid
in theories with less supersymmetry,
such as QCD.

The most compact expressions for these amplitudes
are obtained with the help of two very efficient notational
devices:  color ordering and the spinor helicity
notation (see for example \DixonWI\ for a review).
Color ordering means that we write the total
$n$-gluon amplitude as
a sum over non-cyclic permutations $\rho$
of the $n$ external legs
\eqn\aaa{
{\cal{A}}_n = \sum_\rho \Tr(T^{a_{\rho(1)}} \cdots
T^{a_{\rho(n)}})\,\widehat{A}(\rho(1),\ldots,\rho(n)),
}
where $T^a$ are generators of the gauge group in the adjoint
representation.
The color-ordered amplitude $\widehat{A}$ is invariant under
cyclic permutations of the external legs and has all of
the gauge group structure stripped away.
We can therefore study  $\widehat{A}$ without needing to specify
any particular gauge group.

The spinor helicity notation relies on the fact that any null
vector $p_\mu$ can be decomposed as
\eqn\aaa{
p_{a \dot{a}} = p_\mu \sigma^\mu_{a \dot{a}}=\lambda_a
\wt_{\dot{a}}
}
into a pair of (commuting) spinors of opposite chirality\foot{In
Lorentzian signature, $\wt$ and $\lambda$ are related
by complex conjugation.
In signature $--++$, $\lambda$ and $\wt$ 
may be chosen to be independent real variables, and we will
do so in this paper since this
signature is the one for which the connection to the
string theory on twistor space is most straightforward.
On the YM side this unusual
signature introduces no real difficulty since
tree-level amplitudes are easily continued to
Lorentzian signature. At loop level the situation is less clear.}.
Furthermore, for any chosen pair $(\lambda_a, \wt_{\dot{a}})$,
it is possible to construct polarization vectors
$\epsilon_\pm^\mu$ of either positive or negative helicity,
which are each unique up to gauge transformations.
We use the epsilon tensor to raise and lower the $a$ and $
\dot{a}$ indices, and we introduce the inner products
\eqn\aaa{
\langle i, j \rangle \equiv
\langle \lambda_i, \lambda_j \rangle = \epsilon_{ab}
\lambda_i^a \lambda_j^b, \qquad 
[ i,j] \equiv [\wt_i,\wt_j] = \epsilon_{\dot{a} \dot{b}}
\wt_i^{\dot{a}} \wt_i^{\dot{b}}.
}

Scattering amplitudes are conveniently expressed not
as a function $\widehat{A}(p_\mu, \epsilon_\mu)$ of the
momenta and polarizations of the $n$ particles, but rather
as a function $\widehat{A}(\lambda_a, \wt_{\dot{a}})$ of
the spinors, with
the particle helicities specified.

Amplitudes in which all or all but one of the
$n$ gluons have the same helicity vanish.
The first non-trivial case, in which $n-2$
gluons have positive helicity and two gluons have negative
helicity, is called the maximally helicity violating (MHV)
amplitude.  The MHV amplitude for $n$ gluons in QCD is
given by the Parke-Taylor formula 
\refs{\ParkeGB,\ManganoXK}, whose ${\cal{N}} = 4$ supersymmetric
generalization may be written as \NairBQ
\eqn\mhv{
\widehat{A}_{\rm MHV}(\lambda,
\wt, \eta)= i g^{n-2} (2 \pi)^4 \delta^4
\left(\sum_{i=1}^n
\wt_i^{\dot{a}} \lambda_i^a\right) \delta^8 \left(
\sum_{i=1}^n \lambda_i^a \eta_{iA} \right) \prod_{i=1}^n
{1 \over \langle i,i+1 \rangle},
}
where $\eta_A$, $A=1,2,3,4$ are superspace coordinates
and $n+1 \cong 1$ is understood.

An amplitude $\widehat{A}(\lambda,\wt,\eta)$ written in
physical space coordinates can be expressed
in twistor space variables
by Fourier transform:
\eqn\fourier{
\widetilde{A}(\lambda,\mu,\psi)=
\int {d^{2n} \wt \over (2 \pi)^{2n}}\,d^{4n} \eta
\,\exp \left[i\sum_{i=1}^n \left({ [\mu_i, \wt_i] + \psi_{i}^A
\eta_{iA}}\right)\right] \widehat{A}(\lambda,\wt,\eta).
}
As reviewed in \WittenNN, YM
scattering amplitudes are always homogeneous
of degree zero
 in the
variables $\lambda$, $\mu$, and $\psi$, so an
amplitude $\widetilde{A}(\lambda,\mu,\psi)$ may be viewed
as a function not on $\C^{4|4}$ but on $\P^{3|4}$, which is
the (super-) twistor space of (super-) Minkowski space.
This space has homogeneous coordinates $(z^I, \psi^A)$,
$I=0,\ldots,3$, $A=1,\ldots,4$ which are identified according
to
\eqn\aa{
(z^I, \psi^A) \cong (t z^I, t \psi^A)
}
for any non-zero complex number $t$.
For the present application we decompose
the bosonic coordinates into $z^I = (\lambda^1, \lambda^2,
\mu^1, \mu^2)$. 

In \WittenNN\ it was conjectured that
in twistor space,
the $n$-particle scattering amplitudes with $q$ negative
helicity and $n - q$ positive helicity gluons
are supported on curves in $\P^{3|4}$
of degree
\eqn\degree{
d=q-1+l,
}
and genus
\eqn\genus{
g  \le l
}
where $l$ 
is the number of YM loops.  

For the MHV amplitudes at tree level, it is easily shown
by evaluating the Fourier transform of \mhv\ that the
amplitude is supported on curves of degree 1 
in twistor space~\WittenNN.
For higher degree the Fourier transform appears very complicated.
Fortunately, we will see in \S 4 that it is much simpler
to take the Fourier transform directly in the topological
B-model before evaluating the more complicated integral over
instanton moduli space.

The simplest non-MHV amplitudes
are those with five gluons of helicities $(++---)$ or $(+-+--)$.
In Lorentzian signature, amplitudes with
$n-2$ negative helicities and $2$ positive helicities
are related to MHV amplitudes by complex conjugation and therefore
are given by the simple formula
\eqn\googly{\eqalign{
&\widehat{A}_{\overline{\rm MHV}}
(\lambda,\wt,\eta)
= i g^{n-2} (2\pi)^4 \delta^4
\left( \sum_{i=1}^n \wt_i^{\dot{a}} \lambda_i^a \right)
\cr
&\qquad\qquad\qquad\qquad\qquad\times
\int d^{4 n} \psi \, \exp\left[{i \sum_{i=1}^n \eta_{iA}
\psi_i^A}\right]
\delta^8 \left( \sum_{i=1}^n \wt_i^a \psi_i^A
\right)
\prod_{i=1}^n {1 \over [i,i+1]}.
}}
In \S 4 we recover the precise formula \googly, for $n=5,$ from
a particular amplitude in the presence of a D1-instanton of degree 2
in the topological B-model.

\newsec{How to Calculate Amplitudes in the B-Model on $\P^{3|4}$}

A string theory describing ${\cal{N}}=4$ SYM
theory without additional states must
clearly be different from the usual critical string theories
which contain infinite towers of massive string states. A few
obvious constraints are:
the theory should have a finite spectrum, it should be globally 
invariant under the four dimensional superconformal group
$SU(2,2|4)$, its target space should be related to the usual 
four dimensional space without the introduction of additional compact
dimensions and, of course, it should reproduce the scattering
amplitudes of the gauge theory.

The proposal put forward in \WittenNN\ is that the string
field theory (SFT) of the open topological B-model whose target space is the
supermanifold $\P^{3|4}$ with D5 and D1 branes describes ${\cal{N}}=4$ SYM.

It is clear that this theory has the first three properties stated above.
In the presence of $N$ D5 branes
and no D1 branes the theory is globally invariant
under the isometry group of $\P^{3|4}$ which is also the 
four dimensional superconformal group
$SU(2,2|4)$.  The bosonic part, $\P^3$, of the supermanifold
is identified with the twistor space of the four dimensional Minkowski
space. The spectrum of physical states was 
analyzed \WittenNN\ along the lines of \WittenCS\ with the result
that the physical states form the ${\cal{N}}=4$ $SU(N)$ SYM multiplet.
The classical
equations of motion are, however, those of the self-dual YM
theory and not those of the full ${\cal{N}}=4$ theory.

The conjecture of \WittenNN\ is that the scattering amplitudes of the
full ${\cal{N}} = 4$ gauge theory are recovered by including D1 branes
in this SFT.
Counting of a particular $U(1)$ charge
violation and fermionic zero modes fixes the properties of the 1-brane
contributing to an amplitude with fixed helicities and number of loops
to the values listed in \degree\ and \genus.

The introduction of an instantonic D1 brane leads to additional
states localized on it. The 1-5 and 5-1 strings each contribute a
single physical state 
$\alpha$ and $\beta$, respectively, in the fundamental
representation of $SU(N)$. It was argued in \WittenNN\ that these 
fields should have fermionic statistics, opposite to the naive expectations. 
The 1-1 strings contribute a $U(1)$ gauge field. The action
for the $\alpha$ and $\beta$
fields follows from the standard SFT action:
\eqn\sft{
S= \int_{{\cal{C}}}\,dz\,\alpha({\bar\partial} + {\cal A})\beta.
}
The integral is taken over the worldvolume of the D1-string, which wraps
some holomorphic curve ${\cal{C}}$ sitting inside $\P^{3|4}$.
Here ${\cal A}$ is the 5-5 string field and the coupling follows 
from $SU(N)$ gauge invariance. The scattering amplitudes are then computed in 
terms of correlation functions of the currents
$J(z)=\beta\alpha\,dz$ while treating ${\cal A}$ as a background field.

The introduction of a fixed D1 brane breaks most of the isometries of
$\P^{3|4}$ and thus the resulting amplitudes cannot be invariant under
four dimensional superconformal transformations. This apparent problem
can be easily fixed by integrating over all possible configurations of the
D1 brane, that is, over all
possible choices of ${\cal{C}}$ in 
\sft\ with genus and degree determined by \degree\ and \genus.

For genus zero, the moduli space of curves of degree $d$
in $\P^{3|4}$
is most efficiently described in terms of degree $d$
maps from $\P^1$ into $\P^{3|4}$. 
The embedding map as a function of the coordinate
$\sigma$ on $\P^1$ can be written in terms of
$4(d+1)$  bosonic parameters and
$4(d+1)$ fermionic parameters as
\eqn\aaa{
z^I=P^I(\sigma) = \sum_{k=0}^d a_k^I \sigma^k, \qquad
\psi^A=G^A(\sigma) = \sum_{k=0}^d \beta_k^A \sigma^k,
}
where 
$z^I = (z^0,z^1,z^2,z^3) =
(\lambda^1,\lambda^2,\mu^1,\mu^2)$ are the bosonic
coordinates on $\P^{3|4}$, and $\psi^A$, $A=1,2,3,4$
are the fermionic coordinates.
In this language, the integral over the
moduli space of such curves becomes an integral
over all $a_k^I$ and $\beta_k^A$ while dividing out
by the GL(2) symmetry which acts in the obvious way
on  $\sigma_i$ and (nonlinearly) on
$a$ and $\beta$.

Combining these ingredients leads to the following 
master formula for the tree-level contribution to $n$-gluon scattering
from instantons of degree $d$ (relevant when there are $d+1$ negative 
helicity gluons).
\eqn\main{
B(\lambda,\mu,\psi)
=\int { d^{4d+ 4} a\ d^{4d+4} \beta \ d^n \sigma
\over {\rm vol}(GL(2))}
{\cal J}\,
\prod_{i=1}^n
\delta^3 \left( {z_i^I \over z_i^J }
- { P^I(\sigma_i)
\over
P^J(\sigma_i)}\right)
\delta^4\left( {\psi_i^A \over z_i^J}
-  { G^A(\sigma_i)
\over
P^J(\sigma_i)}
\right).
}
The $3$-dimensional delta function is
taken over $I \ne J$, where the choice of
$J$ is easily seen to be arbitrary.
The final ingredient is the free fermion
correlator
\eqn\aaa{
{\cal J} = \prod_{i=1}^n { 1 \over
\sigma_i - \sigma_{i+1}}.
}
Note that the coordinates $z^I$, $\psi^A$ and $\sigma_i$, as
well as the moduli $a^I_k$ and  $\beta^A_k$ are all complex
variables,
so in writing the formula \main\ we must specify an
integration contour.
In signature $--++$ it is natural to choose the naive
contour where all variables lie on the real axis \WittenNN.

According to the proposal of \WittenNN, \main\ gives
a contribution to the tree-level YM amplitudes Fourier 
transformed to twistor space via \fourier.
In \WittenNN\ the possibility was considered
that there might be additional contributions coming from separated
instantons of lower degree.  In the case we consider next, $n=5$
and $d=2$, we find agreement with YM theory without needing
such contributions.

\newsec{The B-Model Calculation for $n=5$, $d=2$}

In this section we evaluate the B-model amplitude \main\ for
the case $n=5$, $d=2$, which is relevant to the scattering of
3 negative and 2 positive helicity gluons in YM theory.
It is reasonably straightforward to
evaluate the integral \main\ directly in this case\foot{Counting
the bosonic delta functions reveals that the result
will be proportional to two delta functions.  A not so difficult
calculation reveals that they are $\delta(K_{1234})
\delta(K_{1235})$, where $K$ is the object in $\WittenNN$ which
expresses the constraint
that four points lie on a plane $\P^2$ inside $\P^3$.
Note that any five points which lie on a common plane
automatically lie on a degree 2 curve.
It is not clear how to directly Fourier transform (a complicated
function times) $\delta(K_{1234}) \delta(K_{1235})$ back
to physical space.
}.
However,  the quantity we are interested in comparing to
a YM amplitude is not $B$, but its Fourier transform $\widetilde{B}$.
It appears quite intractable to first calculate $B$
and then take the Fourier transform, so we will
proceed by taking the Fourier transform first, before evaluating
the integral over moduli space.

For the next few steps we forget about the fermionic
factor, restoring it at the end of the calculation.
To simplify the already complicated notation, we define the rescaled
variables $\lambda_i^2\rightarrow \lambda_i^2/\lambda^1_i$ and
$\mu_i^{\dot{a}} \rightarrow \mu_i^{\dot{a}}/\lambda^1_i$. The
dependence on $\lambda^1$ can be easily restored at the end, by the
inverse transformation. Fourier transforming the original variable
$\mu^{\dot{a}} \to \wt^{\dot{a}}$ then gives
\eqn\aaa{
\eqalign{
\widetilde{B}(\lambda, \wt) =
&\int { d^{12} a\,d^5 \sigma
\over {\rm vol}(GL(2))}
J_0
\left[
\prod_{i=1}^5
\delta \left( {\lambda_i^2 }
- { P^1(\sigma_i) \over P^0(\sigma_i) } \right)
\right]
\exp \left[ i \sum_{i=1}^5
\sum_{k=0}^2
{\epsilon_{\dot{a} \dot{b}}
\wt_i^{\dot{a}} a_k^{\dot{b}} \sigma_i^k  \over P^0(\sigma_i) }
\right],
}}
where $\wt$ is related to the dual of $\mu$ by the
dual rescaling $\wt_i^{\dot{a}} \rightarrow
\wt_i^{\dot{a}} \lambda_i^1$ and we can
absorb the associated factor of $\prod (\lambda_i^1)^2$ coming
from the measure of the Fourier transform
into
\eqn\aaa{
J_0 = \prod_{i=1}^5 { (\lambda_i^1)^2
\over \sigma_i-\sigma_{i+1}}.
}

The first step is to fix the GL(2) symmetry by setting
the variables
$a_0^0$, $\sigma_1$, $\sigma_2$ and $\sigma_3$ to some arbitrary
values at the cost of introducing the Jacobian
\eqn\aaa{
J_1 = a_0^0 V_{123}, \qquad V_{123} \equiv (\sigma_1 - \sigma_2)(\sigma_2
- \sigma_3)(\sigma_3 - \sigma_1).
}
The integral over the six $a_k^{\dot{a}}$ moduli is trivial and gives
\eqn\six{
\widetilde{B}=
\int  d^{2}a\,d^{3}b\,d\sigma_4\,d\sigma_5
J_0 J_1
\left[
 \prod_{i=1}^5
\delta \left( {\lambda_i^2 }
- { B_i \over A_i } \right)
\right]
\prod_{k=0}^2
\delta^{2} \left(
\sum_{i=1}^5 { \wt_i^{\dot{a}}  \sigma_i^k \over
A_i} \right).
}
Here we have parametrized the remaining bosonic moduli
by $a_k$ (with $a_0=a_0^0$ unintegrated) and $b_k$, with
\eqn\aaa{
A_i = \sum_{k=0}^2 a_k \sigma_i^k, \qquad
B_i = \sum_{k=0}^2 b_k \sigma_i^k.
}
The next step is to make use of the remarkable identity
\eqn\remarkable{
\eqalign{
&\left[ \prod_{i=1}^5
\delta \left( {\lambda_i^2 }
- { B_i \over A_i } \right)
\right]
\prod_{k=0}^2
\delta^{2} \left(
\sum_{i=1}^5 { \wt_i^{\dot{a}}  \sigma_i^k \over
A_i} \right)
\cr
&
\quad= J_2
\delta^2 (
\sum_{i=1}^5 \wt_i^{\dot{a}}  )
\delta^2 (
\sum_{i=1}^5 \wt_i^{\dot{a}} \lambda_i^2 )
\left[ \prod_{i=1}^3
\delta \left( {\lambda_i^2 }
- { B_i \over A_i } \right)\right]
\prod_{k=1}^2
\delta (S_4^k - \sigma_5 S_4^{k-1})
\delta (S_5^k - \sigma_4 S_5^{k-1}),
}}
where
\eqn\sdef{
J_2 = A_4 A_5 [4,5]^3, \qquad
S_j^k = \sum_{i=1}^3 { [i,j] \over A_i} \sigma_i^k.
}
This identity has a number of useful consequences.
Notably, the first two delta functions combine into
overall delta function of momentum conservation
$\delta^4 (p)=\delta^4(\sum_{i=1}^5 \wt_i^{\dot{a}} \lambda_i^{\dot{a}})$
after restoring $\lambda^1_i$ dependence.
Moreover, the $\sigma_4$
and $\sigma_5$ variables now appear linearly in the delta functions.
These, as well as the three $b$ moduli, can therefore be integrated
out with ease.  The latter give a Jacobian of
\eqn\aaa{
J_3 = { A_1 A_2 A_3 \over V_{123}},
}
and we are left with
\eqn\laststep{
\widetilde{B} =
\delta^4 (p)
\int d^2 a \ J_0 J_1 J_2 J_3\ \delta(
S_4^0 S_4^2 - (S_4^1)^2)
 \delta(
S_5^0 S_5^2 - (S_5^1)^2).
}

It is at this stage that the most remarkable feature of the
identity \remarkable\ emerges.
After substituting the definition \sdef\ for $S_j^k$ into
\laststep, the remaining two delta functions turn out to be
linear in the remaining moduli $a_1$, $a_2$!
Integrating them out gives one final Jacobian,
\eqn\aaa{
J_4 = { A_1 (A_2 A_3)^2
\over
J_1 S_4^0 S_5^0([4,2][5,3](\sigma_2-\sigma_5)^2(\sigma_3-\sigma_4)^2-
[4,3][5,2](\sigma_2-\sigma_4)^2(\sigma_3-\sigma_5)^2)}.
}
Now we assemble all of the Jacobians that have piled
up along the way and plug in the values of the moduli
and $\sigma_4$, $\sigma_5$ set by the various delta functions.
In this way we obtain
\eqn\boson{
\widetilde{B} = \delta^4 (p)
{ 
}
\left[
{ [2,1] [3,1] [2,4] [3,4] [2,5] [3,5] a_0^3 (\sigma_2-\sigma_3)^6
\over [4,1]^2 [5,1]^2[3,2]^2 (V_{123})^3
}
\right]^4
\prod_{i=1}^5
{(\lambda_i^1)^2 \over [i, i+1]}.
}
There is significant ambiguity in writing this formula
since $a_0$ is a completely free parameter---one could choose
$a_0$ such that the whole quantity in brackets is 1, for example.
In writing \boson\ we have chosen $a_0$ such that $A_1$, when evaluated
on the solution of all the delta functions, is 
independent of
$\sigma_1$, $\sigma_2$ and $\sigma_3$.
This ambiguity will cancel against the fermionic determinant to
be calculated next.

The final step is to evaluate the fermionic contribution to
the amplitude,
\eqn\ferm{
F \equiv
\int d^{12} \beta \prod_{i=1}^5
\delta^4 \left( \psi_i^A - \sum_{k=0}^2
{ \beta_k^A \sigma_i^k \over A_i}\right).
}
To accomplish this we use a simple analogue of \remarkable\ which
lets us pull out the super-momentum conservation constraint:
\eqn\fdet{
\prod_{i=1}^5
\delta^4 \left( \psi_i^A - \sum_{k=0}^2
{ \beta_k^A \sigma_i^k \over A_i}\right)
= {1 \over [4,5]^4} \delta^8
\left( \sum_{i=1}^5 \wt_i^{\dot{a}} \psi_i^A\right)
\prod_{i=1}^3 \delta^4 \left( \psi_i^A - \sum_{k=0}^2
{ \beta_k^A \sigma_i^k \over A_i}\right).
}
Inserting \fdet\ into \ferm\ and doing
the $\beta$ integrals immediately gives the fermionic
determinant
\eqn\aaa{
F = 
\left[
{V_{123}
\over A_1 A_2 A_3 [4,5]}
\right]^4\delta^8
\left( \sum_{i=1}^5 \wt_i^{\dot{a}} \psi_i^A\right)
}
The quantities $A_i$ are determined in terms of the $\wt^{\dot{a}}_i$
through the bosonic delta functions, all of which we have already
demonstrated how to solve.
Substituting the solutions gives the final expression
\eqn\fermion{
F = 
\delta^8\left( \sum_{i=1}^5 \wt_i^{\dot{a}} \psi_i^A \right)
\left[
{ [4,1]^2 [5,1]^2 [3,2]^2 (V_{123})^3
\over [2,1] [3,1] [2,4] [3,4] [2,5] [3,5] a_0^3 (\sigma_2-\sigma_3)^6}
\right]^4.
}

Combining \boson\ and \fermion\ and restoring the $\lambda_i^1$
dependence by rescaling $\lambda_i^2$, $\wt_i^{\dot{a}}$
and $\psi^A$ as explained in the beginning of this section 
yields the full B-model amplitude
\eqn\aaa{
\widetilde{B}(\lambda,
\wt,\psi)= \delta^4\left( \sum_{i=1}^5 \wt_i^{\dot{a}} \lambda_i^a
\right) \delta^8 \left( \sum_{i=1}^5 \wt_i^{\dot{a}}
\psi_i^A \right) {1 \over [1,2] [2,3][3,4][4,5] [5,1]},
}
in agreement with \googly\ after the necessary fermionic Fourier transform.

\bigbreak\bigskip\bigskip\centerline{\bf Acknowledgments}\nobreak

We have benefited from helpful discussions
with D. Gross, C. Herzog, D. Kosower, W. Siegel and E. Witten.
This work was supported in part by the National Science Foundation
 under Grants PHY99-07949 
(MS, AV) and PHY00-98395 (RR), as well as by the DOE under Grant
No.~91ER40618 (RR).  Any opinions, findings, and conclusions or
recommendations expressed in this material are those of the authors
and do not necessarily reflect the views of the National Science
Foundation.

\listrefs

\end